\documentclass{vienna-conf2019}
\usepackage{graphicx}
\usepackage{hyperref}
\usepackage[]{natbib}  
\usepackage{epstopdf}

\def\BibTeX{{\rm B\kern-.05em{\sc i\kern-.025em b}\kern-.08em
    T\kern-.1667em\lower.7ex\hbox{E}\kern-.125emX}}
\bibpunct{(}{)}{;}{a}{}{,}  

\begin{document}

\TitreGlobal{Stars and their variability observed from space}


\title{Flares of M-stars in  Upper Scorpius region and flares and CMEs of the active M-star AD\,Leo}

\runningtitle{Flares and CMEs of young M-stars}

\author{E. W. Guenther$^{1}$, D. W\"ockel}\address{Th\"uringer Landessterwarte Tautenburg, Sternwarte 5, 
             07778 Tautenburg, Germany}
\author{P. Muheki}\address{Mbarara University of Science and Technology,
P.O Box 1410, Mbarara, Uganda.}




\setcounter{page}{237}


\maketitle


\begin{abstract}
Using the Kepler K2 data, we studied the flare-activity of young K- and M-stars in the Upper Sco region
and found that they have 10000 to 80000 times as many super-flares  with $(E\geq 5\,10^{34} erg)$  
than solar like stars. The power-law index for flares is $N/dE\sim E^{-1.2}$  for K-stars, 
$N/dE\sim E^{-1.4}$ for early M-star and dN/dE~E-1.3 for late M-stars, which is about the 
same as that of the Sun. We also observed the active M-star AD Leo spectroscopy for
222 hours, detected 22 flares but no coronal mass ejections. 
\end{abstract}



\section{Low-mass planets, flares and CMEs of M-stars}

Searching for planets of M-stars has recently become fashionable, because it is relatively easy to detect low mass planets in the so-called habitable 
zone (HZ) of these stars. However, planets in the HZ of M-stars are exposed to the radiation from flares and to Coronal Mass-Ejections (CMEs).  
A  large flare and CME-rate could be critical, because the X-ray and UV-radiation from flares together with the CMEs might erode planetary 
atmospheres. In less extreme cases flares and CMEs still affect the photochemistry of the atmospheres. Particularly important are 
the first few 100 Myrs, because this is the main erosion phase of planetary atmospheres. In here we present the first result of our
study of flares and CMEs on young M-stars. 

\section{Flares of M-stars in  Upper Scorpius}

\begin{figure}[ht!]
 \centering
 \includegraphics[width=0.50\textwidth,clip,angle=-90.0]{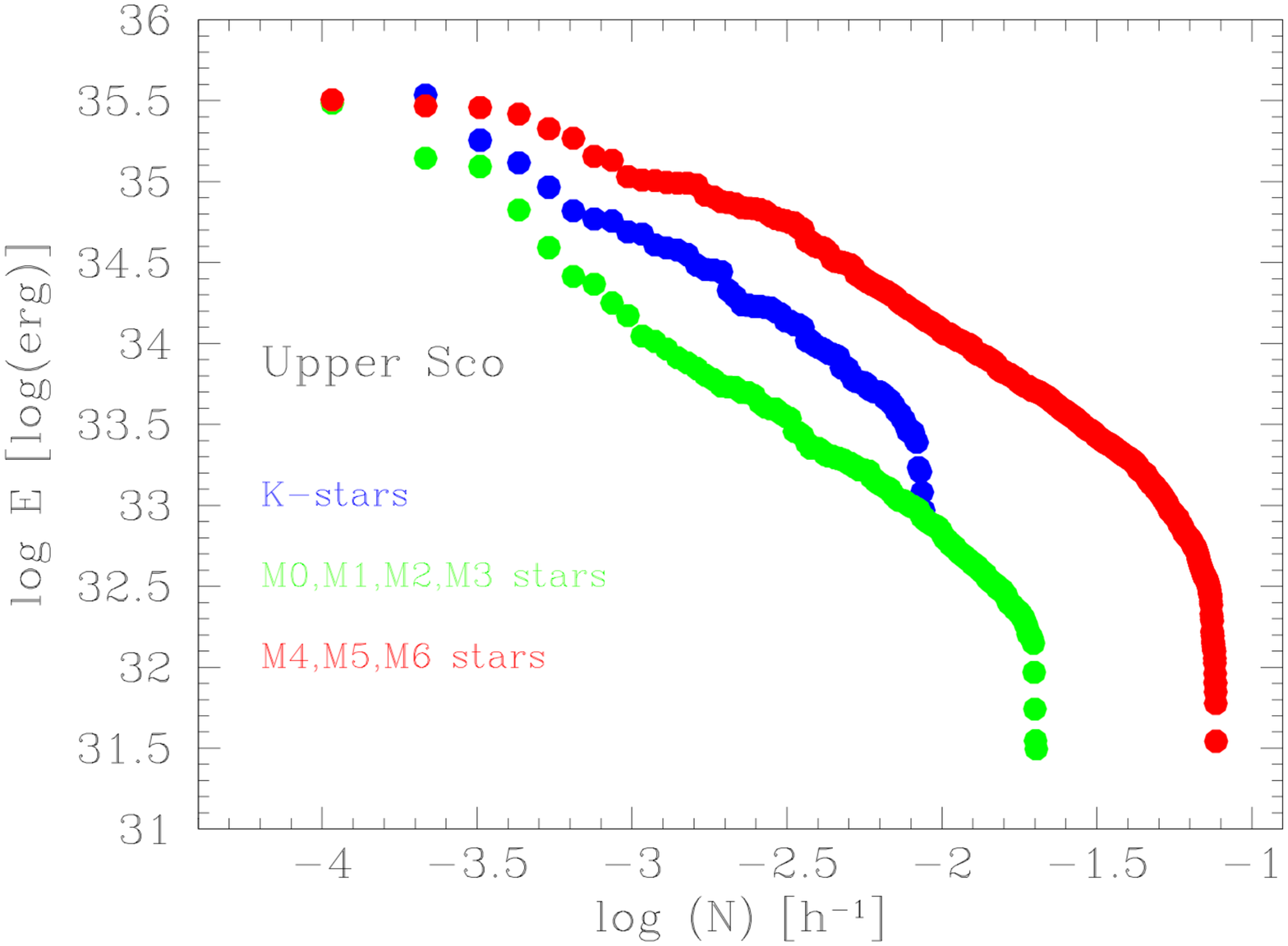}      
  \caption{Cumulative frequency digram for K- and M-stars in Upper Sco.}
   \label{Guenther:fig1}
\end{figure}

Using light-curves obtained in the Kepler K2-mission, we study the flare-activity of M-stars in the  Upper Scorpius OB association which has an age of 
6-11 Myr (Fang et al. 2017). At this age planets have just formed and should be in the main erosion phase of their atmospheres. About 100 M-stars 
in Upper Sco have been observed but we focus here only on those 41 stars of which we confirm as
being members using VLT-FLAMES spectra. Of these 5 are K-stars, 17 early M-stars (M0,M1,M2,M3) and 19 late M-stars (M4,M5,M6). 
Particularly remarkable is the M1-star 2M161115.3-175721 in which we detect 138 flares 
during the 78 days of monitoring. The radius of this stars is 1.30 $R_{sun}$, its mass of 0.85 $M_{sun}$. 
2M161115.3-175721 will  eventually become a late G, or early K-star. Its X-ray brightness is log(Lx)=30.2 $\rm log (erg s^{-1})$ and 
its rotation period of 6.03 days. For comparison: The Sun has log(Lx) 26.4-27.7 $\rm log (erg s^{-1})$, and solar-like stars in the 
Pleiades 29.1-29.6  $\rm log (erg s^{-1})$
(Giardino et al. 2008).  The X-ray flux thus is 300-6000 times larger than that of the Sun. The largest flare
emitted  $3.3\,10^{30}$ erg in the Kepler band (420-900nm), and the smallest 
$10^{27}$ erg.  

Summing up the observing times for all stars in each class, we monitored K-stars for 9273 hours (=1.01 years), 
early M-stars for 31527 hours  (=3.6 years), and late M-stars for 35236 hours (4.0 years). 
We have detected 81 flares in K-stars, 711 in early M-stars and 188 in late M-stars. 
We find that the power-law index of the distribution is $N/dE\sim E^{-1.2}$ for K-stars (log(E)=33.8-35.5), $dN/dE\sim E^{-1.4}$ for early M-stars (log(E)= 32.4-34.2), and $\rm dN/dE\sim E-^{1.3}$ (log(E)33.2-35.0 log(erg)) (Fig.~\ref{Guenther:fig1}). For comparison that Sun has 
$\rm dN/dE \sim E^{-1.5}$ for flares (Crosby et al. 1993), and $dN/dE \sim E^{-1.8}$ for Nanoflares (Aschwanden et al. 2000). 
Flares larger than $5\,10 ^{34}$ erg are usually called super-flares. Compared to solar-like stars, the rate of super-flares is 20000 times 
larger for the K-stars in Upper Sco, 80000 times larger for the early M-stars, and  10000 times larger for the late M-stars (Notsu et al. 2019). 

\section{The frequency of  coronal mass ejections in AD Leo}

Coronal Mass Ejections (CMEs) can be detected in stars as blue-shifted components in the spectrum that have velocities larger than the escape 
velocity of the star. For our study, we selected the M4-star AD Leo which has an age of 20-300 Myr. The escape velocity from this star is 
580 $\rm km\,s^{-1}$. We observed it for 222 hours with the Echelle spectrograph of the 2-m telescope
Alfred-Jensch-Teleskope in Tautenburg and detected 22 flares. The largest of them emitted $2.9\,10^{31}$ erg in H$\alpha$ and 
$1.8\,10^{32}$ erg in $H\beta$. We estimate an XUV-flux to about $2\,10^{33}$ erg for this flare from the H$\alpha$ and 
$H\beta$ fluxes. If M-stars have the same CME/flare ratio as the Sun, we expect to see a
number of CMEs with masses up to $\rm 2-3\,10^{17}\,g$. So far we have not detected a blue-shifted component with 
$V\geq 600 km\,s^{-1}$ but we did observe velocities up to 200 km/s. It thus looks like that the frequency of CMEs on AD Leo is 
lower than expected, unless they escape the detection because of a very high velocity.

\section{Summary and conclusions}

\begin{itemize}
\item Young K- and M-stars have 10000 to 80000 times as many super-flares 
         $(E\geq 5\,10^{34} erg)$  as solar like stars. 
\item This enormous activity will certainly affect the evolution and habitability of  M-star planets. 
\item The power-law index for flares is $N/dE\sim E^{-1.2}$  for K-stars, $N/dE\sim E^{-1.4}$ for 
         early M-star and dN/dE~E-1.3 for late M-stars, which is about the same as that of the Sun. 
\item We observed AD Leo spectroscopically for 222 hours to search for CMEs. 
         In total 22 flares were observed but none of them shows a clear CME signature. 
\end{itemize}

\bibliographystyle{aa}  

\section{References}
\bibliography{aschwanden00} Aschwanden, M.J. et al. 2000, ApJ 535, 1047. \\
\bibliography{crespo06} Crespo-Chac\'on et al. 2006, A\&A 452, 987. \\
\bibliography{crosby93} Crosby, N.B. et al. 1993, SP, 143, 275. \\
\bibliography{fang17} Fang et al. 2017, ApJ 842, 123. \\
\bibliography{giardino08} Giardino et al. 2008, A\&A 490, 113. \\
\bibliography{notsu19} Notsu, Y. et al. 2019, ApJ 876, 58. \\

\end{document}